\documentclass[superscriptaddress,twocolumn]{revtex4-1}
\usepackage{graphicx}                   
\usepackage{hyperref}                   
\usepackage{amsmath,amssymb}            
\usepackage{multirow}		
\usepackage{xcolor}	
\usepackage{amsmath,color}

\graphicspath{{figs/}}					

\DeclareMathAlphabet{\mathpzc}{OT1}{pzc}{m}{it}

\begin{document}

\author{Sina Saber}
\affiliation{Department of Physics, University of Tehran, P. O. Box 14395-547, Tehran, Iran}

\author{Abbas Ali Saberi}\email{(corresponding author) ab.saberi@ut.ac.ir}
\affiliation{Department of Physics, University of Tehran, P. O. Box 14395-547, Tehran, Iran}
\affiliation{Institut f\"ur Theoretische
  Physik, Universit\"at zu K\"oln, Z\"ulpicher Str. 77, 50937 K\"oln,
  Germany}

\title{Universal scaling and criticality of extremes in random matrix theory}


\begin{abstract}
We present a random-matrix realization of a two-dimensional percolation model with the occupation probability $p$. We find that the behavior of the model is governed by the two first extreme eigenvalues. While the second extreme eigenvalue resides on the moving edge of the semicircle bulk distribution with an additional semicircle functionality on $p$, the first extreme exhibits a disjoint isolated Gaussian statistics which is responsible for the emergence of a rich finite-size scaling and criticality. Our extensive numerical simulations along with analytical arguments unravel the power-law divergences due to the coalescence of the first two extreme eigenvalues in the thermodynamic limit. We develop a scaling law that provides a universal framework in terms of a set of scaling exponents uncovering the full finite-size scaling behavior of the extreme eigenvalue's fluctuation. Our study may provide a simple practical approach to capture the criticality in complex systems and their inverse problems with a possible extension to the interacting systems.      
\end{abstract}

\maketitle

Random matrix theory (RMT) and percolation theory are considered as two cornerstones of probability theory with numerous applications in statistical physics and description of various complex systems \cite{r1, r2, r3, r4, r5, r6, r7}. There have been also simultaneous contributions of both theories in the past to describe diverse physical problems ranging from the integer quantum Hall effects \cite{r9} to wireless communications \cite{r10}. RMT and percolation offer their own distinct universal classification schemes characterized by universal distribution functions of their extremes and a set of critical exponents, respectively. The emergence of extreme value theory in the dynamics of percolation models has just recently been reported \cite{r11}. However, to our best knowledge, there has been no report if these two theories are mixed. Here we aim to fill this gap by presenting a random matrix realization of a two-dimensional (2D) percolation model, and discuss how a universal scaling framework emerges near the criticality.  

We directly map a percolation problem with occupation probability $p$ on a square lattice of linear size $L$, to a $L\times L$ random matrix $\mathcal{M}'$ where every element $\mathcal{M}'_{ij}$ is equal to $+1$ with probability $p$ or $-1$ with probability $1-p$. A bond percolation model of this kind is known \cite{r6, r7, r8} to exhibit a continuous phase transition at the critical threshold $p_c=1/2$ with a step-like percolation probability function about $p_c$ in the thermodynamic limit, and power-law divergent correlation length $\xi \propto|p-p_c|^{-\nu}$ and mean cluster size $\chi \propto|p-p_c|^{-\gamma}$ with a bunch of characteristic universal exponents $\nu$, $\gamma$, etc. In this letter, we report on a manifestation of such critical concepts in the random matrix realization of a percolation problem. We shall indeed refer to the matrix and percolation ensembles interchangeably. 

In our mapping, the matrix elements
$\mathcal{M}'_{ij}$ are chosen to be either $-1$ or $+1$ (instead of being $0$ or $1$), to satisfy the mean zero condition at $p_c=1/2$ which typically appears in the random matrix literature, and the additional advantage of symmetry $\mathcal{M}'_{ij}\leftrightarrow -\mathcal{M}'_{ij}$ when $p\leftrightarrow 1-p$. Here, the statistics of extreme eigenvalues is of interest. In order to have real eigenvalues, we use symmetrization $\mathcal{M}=(\mathcal{M}'+\mathcal{M}'^T)/2$, where $(\cdot)^T$ denotes the transpose of the matrix. Thus, the real symmetric matrix $\mathcal{M}$ has $\frac{1}{2}L(L+1)$ independent elements $\mathcal{M}_{ij}$ which can take values $-1$, $0$ and $+1$ with probabilities $(1-p)^2$, $2p(1-p)$ and $p^2$, respectively, for $i>j$. The diagonal elements $\mathcal{M}_{ii}$ can be either $-1$ or $+1$ with probabilities $(1-p)$ or $p$, respectively. Every element has the average $\mu=\langle\mathcal{M}_{ij}\rangle=2p-1$. The zero mean condition thus only holds for $p_c=1/2$.

\begin{figure}[t]
	\centering
	\includegraphics[width=0.48\textwidth]{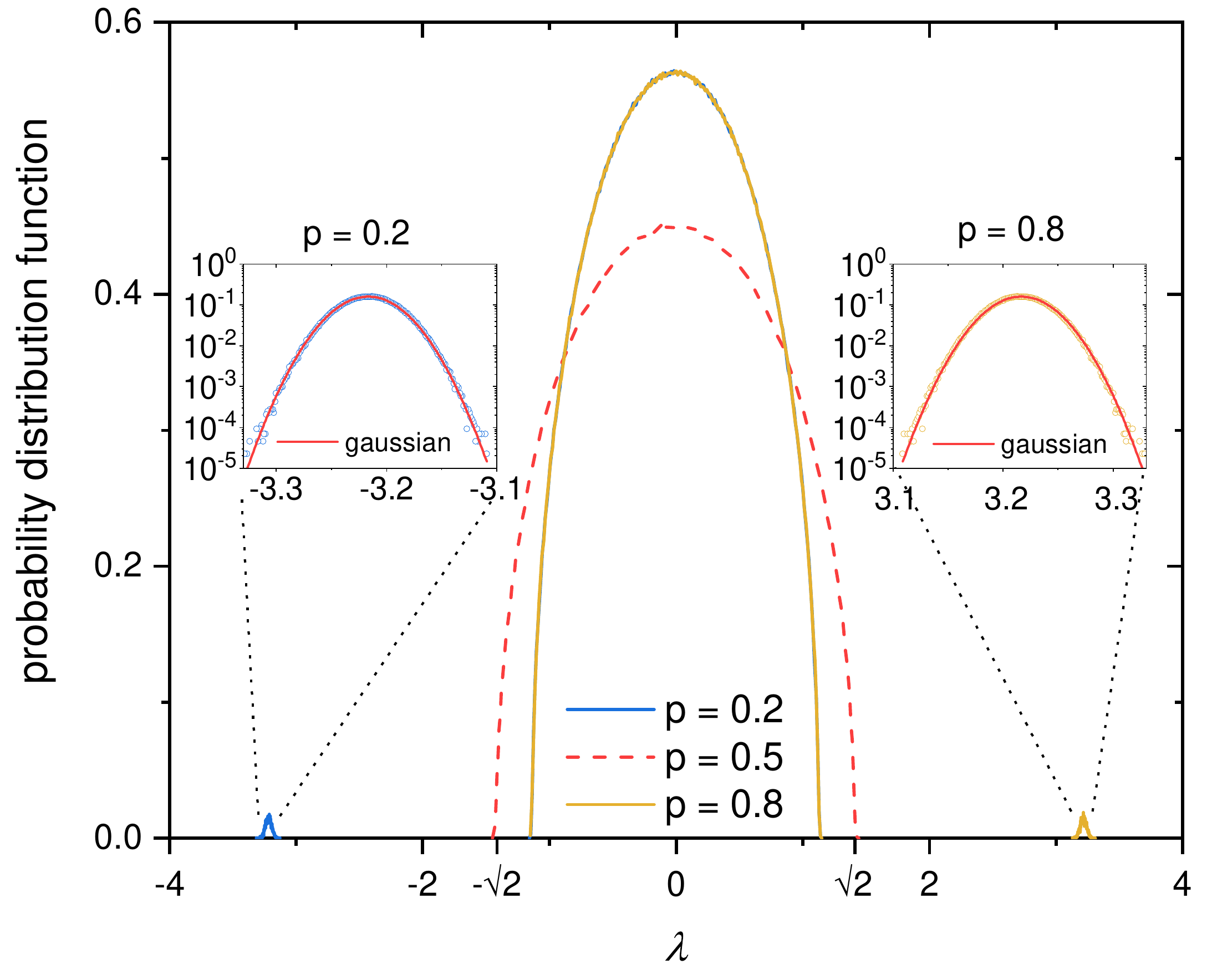}
	\caption{ Eigenvalue distribution of an ensemble of rescaled symmetric random matrices as a realization of a percolation model on a square lattice of size $L=2^{10}$ with occupation probabilities $p=0.2, 0.5$ and $0.8$. For $p=0.5$ (the red dashed line) which satisfies the zero mean condition, reproduces the Wigner's semicircle law with edges exactly located at $\pm\sqrt 2$. The distributions for $p=0.2$ and $0.8$ have two parts: (i) A semicircle law for the bulk eigenvalues whose edges approach zero, and due to the symmetry  $p\leftrightarrow 1-p$, both  collapse on the top of each other for $p=0.2$ and $0.8$, and (ii) A disjoint isolated Gaussian distribution of the extremes (Insets). }
	\label{fig-1}
\end{figure}

The limiting distribution of eigenvalues of various random symmetric matrices is given by Wigner's semicircle law as the size of the matrix approaches infinity, i.e., $L\rightarrow\infty$ \cite{r12, r13}. This law was originally stated for matrices with entries taking the values $\pm 1$ each with probability $1/2$ \cite{r12}, and then it has been generalized to matrices with entries distributed symmetrically about zero and having finite moments of all orders \cite{r13}.\\ Figure \ref{fig-1} presents the probability distribution function of the eigenvalues for an ensemble of rescaled symmetric random matrices $\{\mathcal{M}\}$ as a realization of a percolation model on a square lattice with occupation probabilities $p=0.2, 0.5$ and $0.8$.
The matrices for all $p$'s are rescaled by a factor of $1/\sqrt L$, so that the edges of the semi-circle distribution function of eigenvalues would lie in the interval $[-\sqrt 2, \sqrt 2]$. As shown in Figure \ref{fig-1}, since the zero-mean condition holds for $p=0.5$, the distribution of eigenvalues (the red dashed line) coincides with the Wigner's semicircle law with edges exactly located at $\pm\sqrt 2$. However, the distribution of eigenvalues for $p=0.2$ and $0.8$ consists of two parts: (i) A semicircle part for the bulk eigenvalues whose edges approach zero as $|p-p_c|\rightarrow 1/2$, and due to the symmetry  $p\leftrightarrow 1-p$, both collapse on the top of each other for $p=0.2$ and $0.8$, and (ii) A disjoint isolated distribution of the extremes including the largest (smallest) eigenvalues for $p=0.8$ ($0.2$) with a perfect Gaussian distribution (shown in the Insets of Fig. \ref{fig-1}).

\begin{figure}[t]
	\centering
	\includegraphics[width=0.48\textwidth]{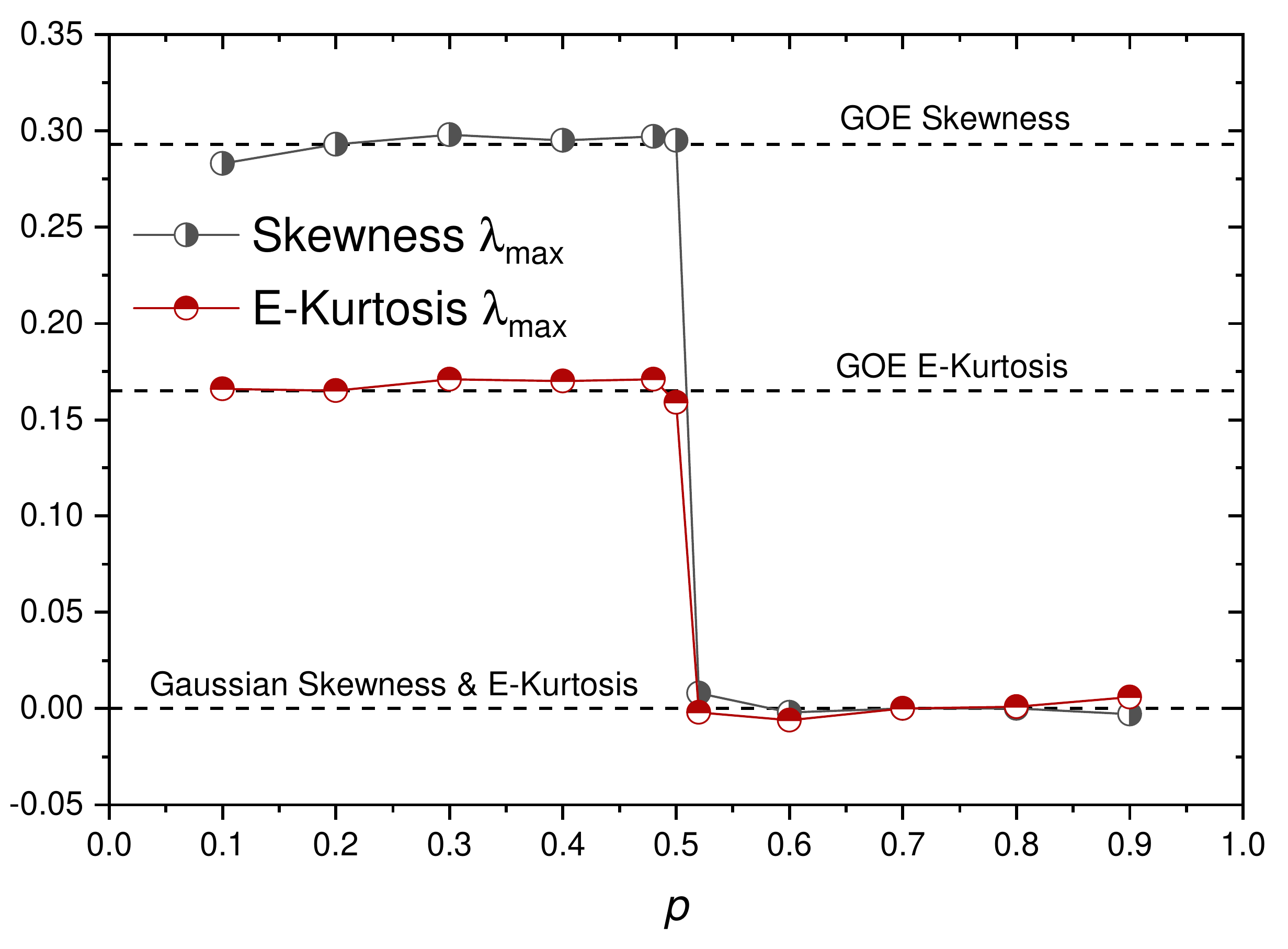}
	\caption{The skewness and the excess kurtosis of the largest eigenvalues $\lambda_{\text{max}}$ estimated from an ensemble of $10^5$ independent percolation realizations on a square lattice of size $L=2^{11}$ as a function of $p$. The horizontal dashed lines indicate the corresponding predictions for the GOE and Gaussian distributions.    }
	\label{fig-2}
\end{figure}

For real symmetric (orthogonal) random matrices, it is known \cite{r14} that the limiting distribution of the largest eigenvalue sitting at the spectral edge is given by the Gaussian Orthogonal Ensemble (GOE) Tracy-Widom distribution \cite{r15} with the characteristic skewness $\sim 0.29346$ and excess kurtosis $\sim 0.16524$. In Figure \ref{fig-2}, we report our estimates for the skewness and the excess kurtosis of the largest eigenvalue $\lambda_{\text{max}}$ as a function of $p$, compared with the corresponding predictions for the GOE and Gaussian distributions (shown by the horizontal dashed lines).  Our results suggest the following: (i) The isolated Gaussian distribution for $p > p_c$ ($p < p_c$) (see Fig. \ref{fig-1}) only includes the first largest (smallest) eigenvalues, and, the second largest (smallest) eigenvalues are sitting at the spectral edges. (ii) The step-like behavior in Fig. \ref{fig-2}, is suggestive of a sharp transition very close to the critical threshold i.e., $p\rightarrow p_c$, when the two Gaussian and GOE distributions merge \cite{r15-1}. We noted that a similar observation has been recently reported in a different context \cite{r16} for the extremal eigenvalues of sparse Erd\H{o}s-R\'enyi graphs. As we will see in the following, this merging in our model is accompanied by some power-law divergences in the statistics of the first largest/smallest extremes which can be described within a remarkable universal scaling framework. 

\begin{figure}[b]
	\centering
	\includegraphics[width=0.4\textwidth]{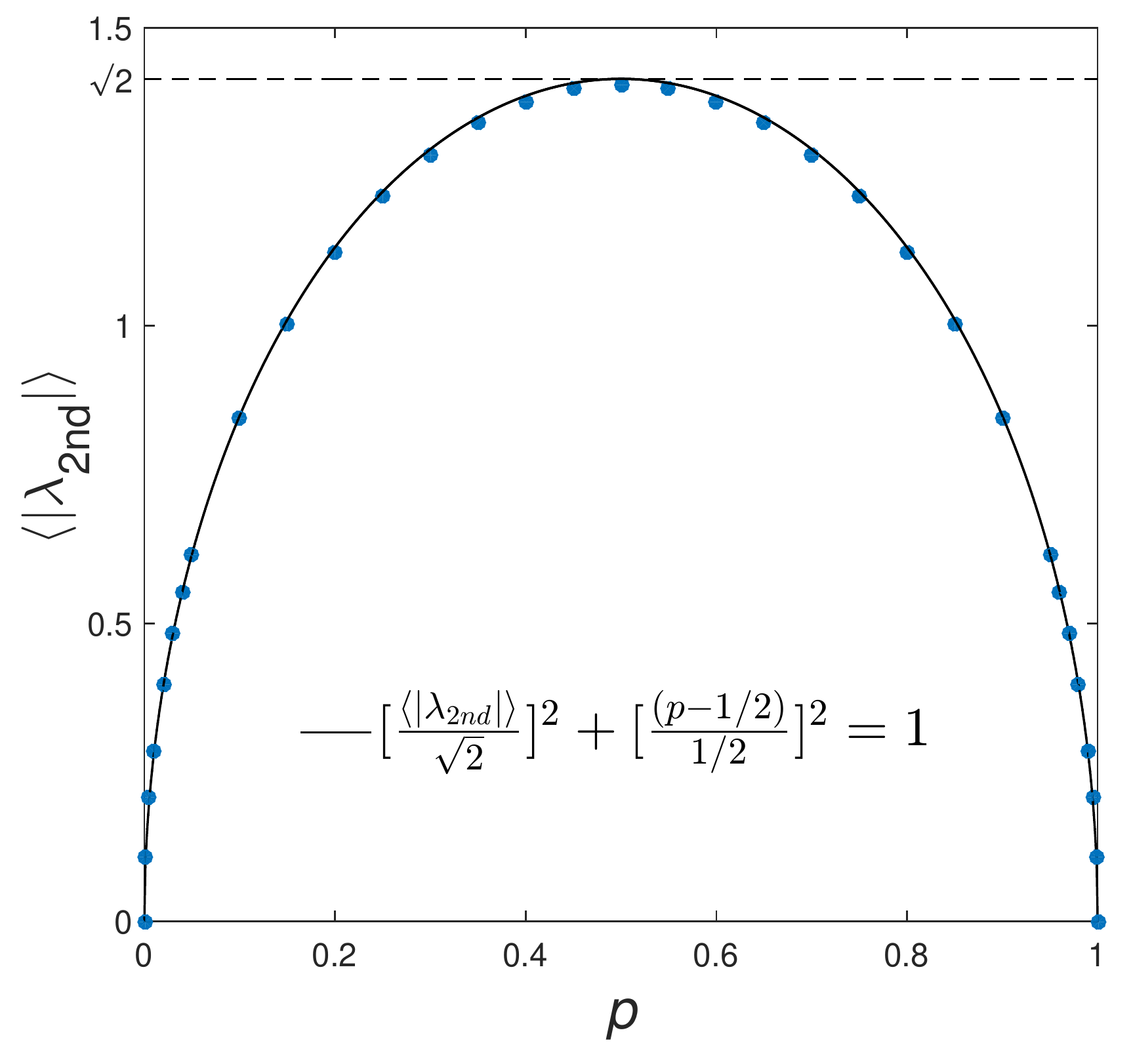}
	\caption{The average absolute value of the second extremes $\langle|\lambda_{\text{2nd}}|\rangle$ in the limit of the infinite system size $1/L\rightarrow 0$ as a function of $p$. The solid line shows a semi-ellipse functionality which well describes our data.  }
	\label{fig-3}
\end{figure}

Let us now investigate how the location of the semicircle edges in Fig. \ref{fig-1} behaves as a function of $p$. To this end, we have measured the average absolute value of the second extremes $\langle|\lambda_{\text{2nd}}|\rangle$ in the limit of the infinite system size $L\rightarrow\infty$ for various choices of $p$. Our computed data are shown by symbols in Fig. \ref{fig-3}. The best fit to our data suggests a perfect agreement with a semi-ellipse functionality shown by the solid line, i.e.,  \begin{equation}\Big[\frac{\langle|\lambda_{\text{2nd}}|\rangle}{\sqrt 2}\Big]^2+\Big[\frac{(p-1/2)}{1/2}\Big]^2=1.\end{equation}
This can be used for the rescaling of the matrix $\mathcal{M}$ by an additional factor $1/\sqrt{1-\langle\mathcal{M}_{ij}\rangle^2}$, to enforce the edges of the semicircles to stay at $\pm\sqrt 2$ for all $p$'s.

\begin{figure}[t]
 	\centering
 	\includegraphics[width=0.49\textwidth]{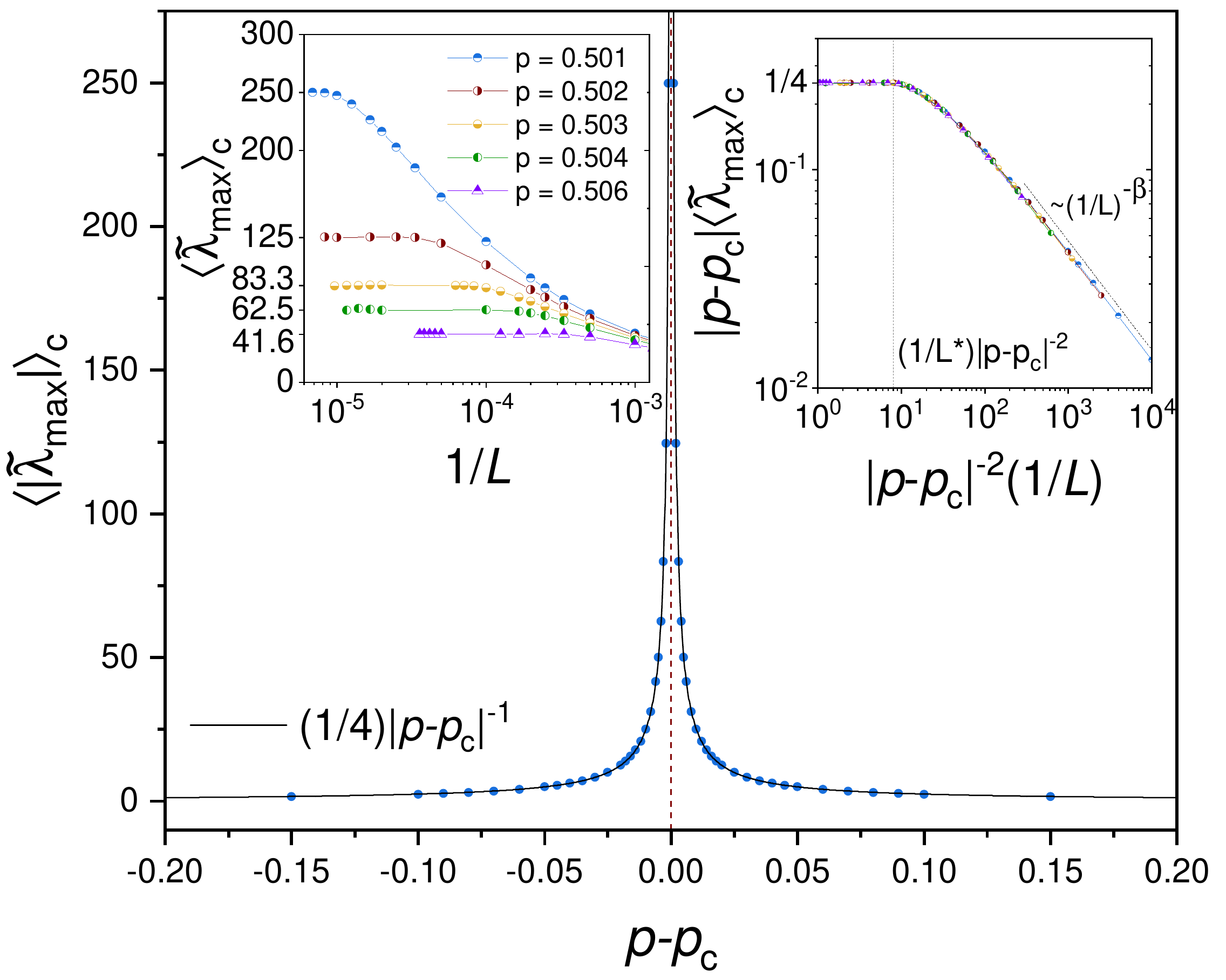}
 	\caption{The scaling law developed for the finite-size behavior of the fluctuations of the first extreme eigenvalues.  Main: Asymptotic estimates of the average fluctuation of the largest absolute eigenvalue versus $p-p_c$. The solid line is the power-law with exponent $-1$ which best describes our data. Left Inset: The average fluctuation of the largest eigenvalue for five occupation probabilities with $p>0.5$ as a function of inverse system size. The average fluctuation increases with system size up to $L^*$ after which it saturates to the asymptotic value. Right Inset: All data that are shown in the left Inset collapse onto a single universal curve when they are suitably normalized according to the scaling formalism proposed in Eq. \ref{F-V-scaling} with $\alpha=1$, $\beta=1/2$ and $z=2$. }
 	\label{fig-4}
 \end{figure}

In the rest of the Letter, we shall focus our attention on the statistical and scaling properties of the isolated extreme eigenvalue and its finite size effects. It is shown in \cite{r17} that for a real symmetric matrix with nonzero mean, there always exists an eigenvalue whose magnitude is greater than or equal to the mean of the sums of the elements in a row of the matrix. This means that the absolute value of the largest eigenvalue in $\mathcal{M}$ is in the order of $|\langle\mathcal{M}_{ij}\rangle| L$, i.e., $\langle|\tilde\lambda_{\text{max}}|\rangle\sim |2p-1| L$ where the tilde denotes for the eigenvalues of 'non-rescaled' matrices (Note  that the largest eigenvalue of a deterministic matrix $\mathcal{M}_{ij}=\mu$ for $i\ge j$, is equal to $\mu L$, and all the other  eigenvalues are equal to zero. Thus the fluctuation of the entries of the matrix each with the variance $\sigma^2$ for $i>j$, changes $\tilde\lambda_{\text{max}}$ only with a normal random variable of expectation $\sigma^2/\mu$ and variance $2\sigma^2$ not growing with the size $L$---see Fig. \ref{fig-1}.). The quantity of interest is indeed the average fluctuation of the extreme eigenvalue i.e., $\langle|\tilde\lambda_{\text{max}}|\rangle_c:=\langle|\tilde\lambda_{\text{max}}|\rangle-|2p-1|L$ in the asymptotic limit $L\rightarrow \infty$.\\ We have carried out extensive Monte Carlo simulations of extremely large system sizes $100\le L\le146400$ of our model in the vicinity of the critical occupancy $p_c=1/2$ involving strong finite size effects, to estimate $\langle|\tilde\lambda_{\text{max}}|\rangle_c$ for every $p$ in the limit $L\rightarrow\infty$. As shown in the left Inset of Figure \ref{fig-4}, the finite-size effects necessarily become essential near the critical point and the convergence of $\langle|\tilde\lambda_{\text{max}}|\rangle_c$ occurs at much larger system size when $p\rightarrow p_c$. Our estimates for the asymptotic $\langle|\tilde\lambda_{\text{max}}|\rangle_c$ as a function of $p$ in the vicinity of $p_c$ are shown in the main panel of Fig. \ref{fig-4}. The best fit to our data suggests the following power-law behavior: 
\begin{equation}\label{-1exponent}
\big\langle|\tilde\lambda_{\text{max}}|\big\rangle_c =\frac{1}{4} |p-p_c|^{-1}.
\end{equation} 
This finding confirms our above mentioned expectation that the average fluctuations about $\mu L$ should asymptotically be given by $\sigma^2/\mu=2p(1-p)/(2p-1)$, which gives rise to the scaling relation given in Eq. (\ref{-1exponent}) as $p(1-p) |p-p_c|^{-1}$ with the identification that the prefactor $p(1-p)\rightarrow 1/4$ when $p\rightarrow p_c=1/2$. 
 
Although Eq. (\ref{-1exponent}) is asymptotically true in the infinite size limit $L\rightarrow\infty$, the left Inset in Figure \ref{fig-4} suggests that the extreme eigenvalue's fluctuation generally depends on the length scale of observation: At a given $p$, the fluctuation grows with system size until a characteristic length scale $L^*$ above which, it saturates to a size-independent value. We are interested in if, in analogy with the roughness growth in non-equilibrium
stochastic surface-growth phenomena, a "Family-Vicsek"-like scaling law provides a universal framework for our observations summarized in Figure \ref{fig-4}. We have examined our hypothesis, and the answer was surprisingly consistent. We find that the following spatio-occupational scale-invariant behavior governs the fluctuations of the first extreme eigenvalue \begin{equation}\label{F-V-law}
\big\langle|\tilde\lambda_{\text{max}}|\big\rangle_c(p, L)=|p-p_c|^{-\alpha} \mathpzc{F}\bigg(\Big(\frac{1}{L}\Big)|p-p_c|^{-z}\bigg),
\end{equation}where the universal function $\mathpzc{F}(x)\rightarrow 1/4$ as $x\rightarrow 0$ (see Eq. (\ref{-1exponent})) and $\mathpzc{F}(x)\sim x^{-\alpha/z}$ as $x\rightarrow \infty$, so that $\langle|\tilde\lambda_{\text{max}}|\rangle_c$ grows with size like $L^{\alpha/z}$ until it saturates to $(1/4)|p-p_c|^{-\alpha}$ when $L\sim |p-p_c|^{-z}$. More specifically, 

\begin{equation}\label{F-V-scaling}
\big\langle|\tilde\lambda_{\text{max}}|\big\rangle_c(p, L)\propto
\begin{cases}
|p-p_c|^{-\alpha} & \text{if $L^*\ll L$}\\
(1/L)^{-\beta} & \text{if $L\ll L^*$},\\
\end{cases}       
\end{equation}
with $L^*$ being a saturation length proportional to $|p-p_c|^{-z}$, and $\alpha$, $\beta$ and $z=\alpha/\beta$ are three positive scaling exponents featuring the universality of the extreme eigenvalue's fluctuation.\\As shown in the right Inset of Fig. \ref{fig-4}, we find that a perfect collapse onto a universal function $\mathcal{F}$ is achieved when the data is appropriately rescaled by the following examined (conjectured) exponents\begin{equation}\nonumber\alpha=1, \hspace{0.8cm} \beta=1/2, \hspace{0.8cm}  z=2.\end{equation} Within the standard Family-Vicsek scaling law in surface-roughness growth, one may note that the same corresponding set of exponents describes the universality of the linear Edward-Wilkinson (EW) equation \cite{r18} (with exactly known exponents $\alpha=(2-d)/2$, $\beta=(2-d)/4$ and $z=2$) in $0+1$ dimensions.

\begin{figure}[t]
	\centering
	\includegraphics[width=0.49\textwidth]{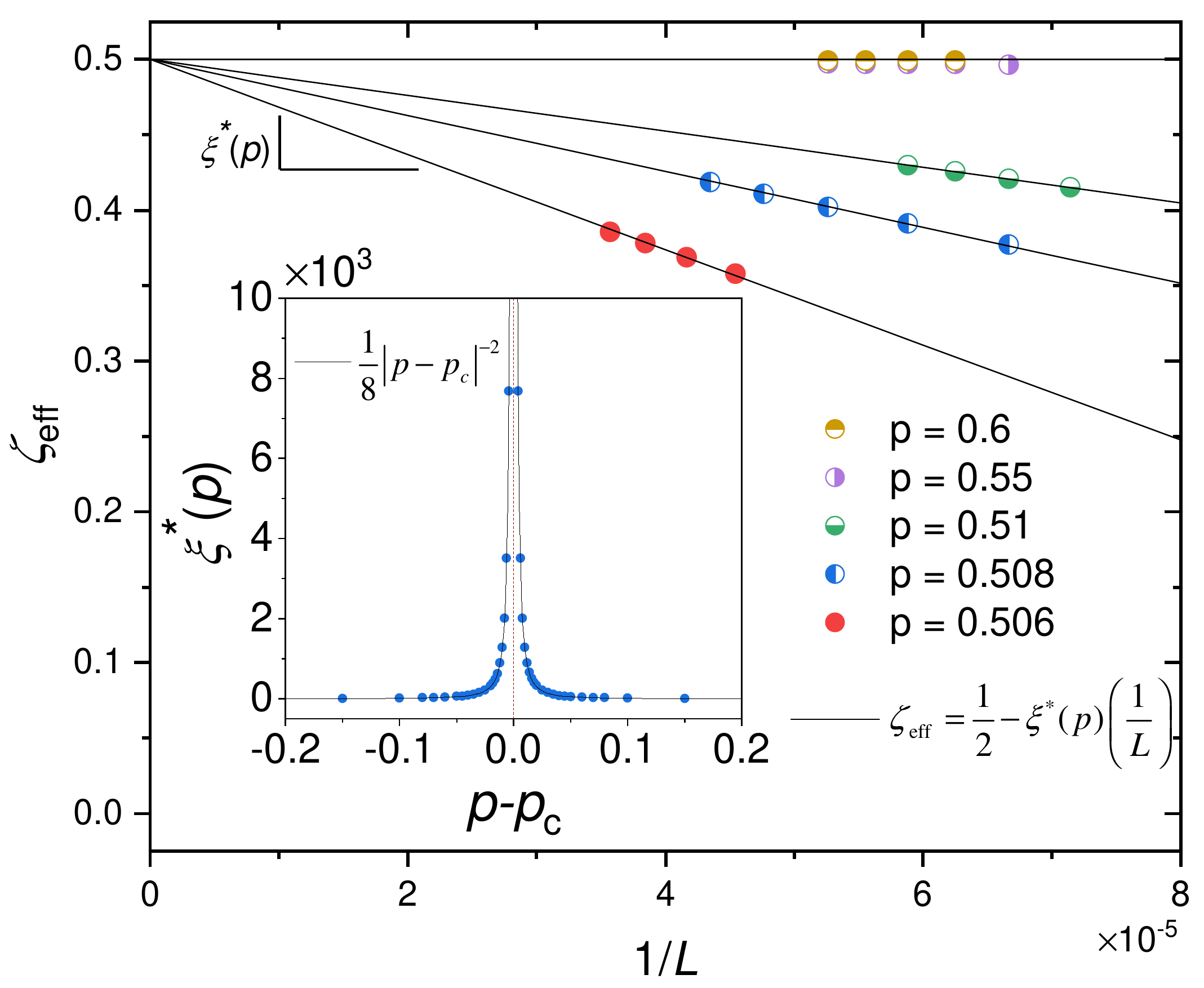}
	\caption{Main: The effective exponent $\zeta_{\text{eff}}$ defined in Eq. (\ref{alpha_eff}) as a function of $1/L$ for different values of occupation probability $p$ in the vicinity of the critical point $p_c$. All data behave linearly with $1/L$ with a negative slope $\xi^*(p)$ and clear extrapolations to $1/2$ in the infinite-size limit (solid lines). Inset: Power-law divergence of $\xi^*(p)$ measured from the data presented in the Main panel. The solid line shows the power-law $\frac{1}{8}|p-p_c|^{-2}$ which best describes our data. }
	\label{fig-5}
\end{figure}

Until now, we have focused on the scaling properties of the 'non-rescaled' average extreme eigenvalue fluctuations around the mean value $|2p-1| L\propto L^1$. This means that the leading size contribution in the 'rescaled' average extreme eigenvalue is proportional to 
$L^{\zeta}$ with $\zeta=1/2$. We are interested in realizing to what extent the exponent $\zeta=1/2$ for the average extreme eigenvalue $\langle|\lambda_{\text{max}}|\rangle$ remains valid when $p\rightarrow p_c$. Without loss of generality, in the following we discuss the supercritical regime $p>p_c$with a straightforward extension to $p<p_c$ regime due to the symmetry $p\leftrightarrow 1-p$.
Our attempt to carefully validate the exponent $\zeta=1/2$, has led to capture the power-law behavior we have already observed for $L^*$ in more detail. \\In order to delicately estimate the exponent $\zeta$ characterizing the dominant size dependence of $\langle\lambda_{\text{max}}\rangle$, we define an effective exponent $\zeta_{\text{eff}}$ which allows to easily monitor the approach to the true asymptotic behavior. For a given occupation probability $p$, $\zeta_{\text{eff}}(L_k)$ is defined by the successive slopes of the line segments connecting two neighboring points of $\big(L_{k-1}, \langle\lambda_{\text{max}}\rangle (L_{k-1})\big)$ and $\big(L_{k}, \langle\lambda_{\text{max}}\rangle (L_{k})\big)$ in which $\langle\lambda_{\text{max}}\rangle (L_{k})$ stands for the average largest eigenvalue for a system size $L_k$, \begin{equation}\label{alpha_eff}
\zeta_{\text{eff}}(L_k)=\frac{\ln\big[\langle\lambda_{\text{max}}\rangle (L_{k})/\langle\lambda_{\text{max}}\rangle (L_{k-1})\big]}{\ln[L_k/L_{k-1}]}.
\end{equation} It is plotted against $1/L$ in Figure \ref{fig-5} for various choices of $p$ to capture their extrapolations in the infinite-size limit. We observe that the effective exponent increases almost linearly with $1/L$ for all $p>p_c$ with a clear extrapolation to $\zeta=1/2$ in the limit $1/L\rightarrow 0$ (the solid lines). 
However, our remarkable observation from Fig. \ref{fig-5} is that in the sufficiently large system-size regime where the effective exponent $\zeta_{\text{eff}}$  increases linearly with the inverse system size, \begin{equation}\label{xi}
\zeta_{\text{eff}}=\frac{1}{2}-\xi^*(p)\bigg(\frac{1}{L}\bigg),
\end{equation} the corresponding slope $\xi^*(p)$ exhibits a power law divergence as the critical threshold $p_c$ is approached (see the Inset of Fig. \ref{fig-5}). Our best fit to our data, provides the following conjecture in perfect agreement with our data \begin{equation}\label{xistar}
\xi^*(p) =\frac{1}{8} |p-p_c|^{-2}.
\end{equation}
The length scale $\xi^*(p)$ should be identical with $L^*\propto|p-p_c|^{-z}$ with $z=2$, as the only relevant length scale in our study. This is independently verified when we have marked $(1/L^*)|p-p_c|^{-2}$ with the vertical dotted line in the right Inset of Fig. \ref{fig-4} where crosses the horizontal axis at $8$. For every $p>p_c$, $\xi^*(p)$ remains finite and the exponent $\zeta=1/2$ is always achieved in the asymptotic limit $L\rightarrow\infty$. Exactly at the critical occupancy $p_c$, the extreme eigenvalues reside on the spectral edges in the finite interval $[-\sqrt{2}, \sqrt{2}]$ which necessarily needs $\zeta=0$.

Our study can be appropriately generalized to construct a random-matrix representation of a 2D bond percolation model described by a sum of two independent random real matrices $\mathcal{M'}^{h}$ and $\mathcal{M'}^{v}$, one for the horizontal and one for the vertical bonds: $\mathcal{M'}=\mathcal{M'}^{h}+i\mathcal{M'}^{v}$, in which $\mathcal{M'}^{h}_{ij}$ and $\mathcal{M'}^{v}_{ij}$ denote as the horizontal and vertical bonds to the right resp. above site ($i, j$). This associates to each open/closed horizontal bond a value $\pm 1$, and to each open/closed vertical bond a complex value $\pm i$. Therefore, each bond configuration corresponds to a matrix $\mathcal{M'}$ with elements $\pm 1\pm i$. Thus 2D bond percolation on the square lattice is associated with an ensemble of asymmetric complex matrices.\\ We have carried out new simulations to analyze the eigenvalue spectrum of an ensemble of such matrices after symmetrization $\mathcal{M}=(\mathcal{M}'+\mathcal{M}'^t)/2$, where $(\cdot)^t$ denotes the conjugate transpose of the matrix. We find that the same scaling laws proposed in Eqs. (\ref{F-V-law}) and  (\ref{F-V-scaling}) with the same scaling exponents $\alpha=1, \beta=1/2$ and $z=2$ well describe the 2D bond percolation model with the only difference that here the corresponding prefactors in Eqs. (\ref{-1exponent}) and (\ref{xistar}) take on exactly twice values, i.e., $\big\langle|\tilde\lambda_{\text{max}}|\big\rangle_c =\frac{1}{2} |p-p_c|^{-1}$, and $\xi^*(p) =\frac{1}{4} |p-p_c|^{-2}$.

To conclude, we have presented a random matrix realization of a 2D percolation model which offers a novel formalism to study complex systems in a rather simple approach that is in all aspects consistent with the concepts in critical phenomena. We find that the behavior of the model is encoded in the fluctuations of the extreme isolated eigenvalues whose divergence signals the criticality in the vicinity of the critical point. We have established a scaling law, which provides a universal framework characterizing the finite-size effects in terms of a set of critical exponents. Self-consistency conditions along with analytical arguments let us making exact conjectures for the exponents and prefactors in perfect agreement with our computations which, however, calls for future mathematical attempts for their rigorous proofs. The random matrix realization of interacting and/or correlated statistical models may provide an inroad for progress towards the understanding of elusive models in two dimensions.

\textit{Acknowledgment.} We would like to thank S.N. Majumdar, H. Spohn, and especially P.L. Ferrari for useful discussions. A.A.S. acknowledges the supports from the Alexander von Humboldt Foundation (DE) and the research council of the University of Tehran. We also thank the High-Performance Computing (HPC) center at the University of Cologne, Germany,
where a part of computations has been carried out.
	

\end{document}